# Quantum circuit for exponentiation of Hamiltonians: an algorithmic description based on tensor products

GERARD FLEURY AND PHILIPPE LACOMME

LIMOS - UMR 6158
Campus des Cézeaux, 1 rue de la Chebarde
TSA 60125 - CS 60026
63178 Aubière cedex France
Email: (gerard.fleury@isima.fr, placomme@isima.fr)

**Abstract**

Exponentiation of Hamiltonians refers to a mathematical operation to a Hamiltonian operator, typically in the form $e^{-i.t.H}$, where $H$ is the Hamiltonian and $t$ is a time parameter. This operation is fundamental in quantum mechanics, particularly to evolve quantum systems over time according to the Schrödinger equation. In quantum algorithms, such as Adiabatic methods and QAOA, exponentiation enables efficient simulation of a system's dynamics. It involves constructing quantum circuits that approximate this exponential operation. When $H = \sum_{p=1}^{n} H_p$, each $H_p$ is defined using the Pauli operator basis, which includes the well-known $X$, $Y$, $Z$ and $Id$ gates, i.e., $H_p = U_1 \otimes U_2 \otimes ... \otimes U_n$ and $U_k \in \{Id, X, Y, Z\}$. In this article, we explore the exponentiation of $H_p$, specifically $e^{-i.t.U_1 \otimes U_2 \otimes ... \otimes U_n}$, by introducing an algorithmic approach. We demonstrate a straightforward and efficient method to construct compact circuits that are easy to implement.

## 1. Introduction

As introduced by (Schrödinger, 1926), the evolution of a quantum-mechanical system's wave function is described by the equation:

$$\frac{\partial.}{\partial t}|\psi(x,t)\rangle = -i.\hbar.H(t).|\psi(x,t)\rangle$$

where $H(t)$ represents the system's Hamiltonian (energy operator), where $\hbar$ is derived from Plank constant and where $|\psi(x,t)\rangle$ are the states vectors. If $H(t)$ is time dependent the solution is given by

$$|\psi_t\rangle = e^{-i.\hbar.\int_0^T H(u).du}.|\psi_0\rangle$$

that can be approximated replacing the integral by a sum of integral on small enough time intervals, to consider that $H$ is time independent on each interval. By consequence the ground state can be estimated using adiabatic based approach considering a Hamiltonian $H$ and an initial state $|\psi_0\rangle$.

A specific solution can be found by considering an approximation of the Schrödinger equation, making use of a particular Hamiltonian $H$ Hamiltonian (involving $Z_i$, and $Z_i.Z_j$ operators). In Adiabatic Quantum Optimization, the system is required to evolve slowly into the ground state of a Hamiltonian $H_D$ often called the "driver Hamiltonian": this Hamiltonian should not commute with the one modeling the combinatorial problem.. The system must then be adiabatically tuned from $H_D$ to $H_P$ through an interpolation involving a parameter $s(t)$ a which smoothly decreases from 1 to 0:

$$H(t) = s(t).H_D + [1 - s(t)].H_P \text{ (Farih et al., 2000).}$$

However, Adiabatic Quantum Optimization and other quantum annealing approaches are limited to small-scale instances due to the slow evolution required between $H_D$ to $H_P$. In 2014 (Farih et al., 2014) introduced a new approach known as the Quantum Approximate Optimization Algorithm (QAOA), which provides a more efficient decomposition scheme. QAOA has garnered significant attention, with notable works by (Farhi and Harrow, 2019), (Yang et al., 2017), (Jiang et al., 2017), (Wecker et al., 2016), and (Wang et al., 2018). In a recent study (Farhi and Harrow, 2019), it was shown that the output distribution produced by QAOA cannot be efficiently approximated by any classical algorithm due to the inherent complexity of the required algorithms. In recent years, QAOA has drawn considerable interest from researchers, as it is seen as a potential candidate for demonstrating quantum supremacy, as emphasized by (Farhi and Harrow, 2019). Quantum supremacy refers to the ability of a quantum circuit to produce results that require an exponentially larger number of classical operations to replicate.

The Hamiltonian $H$ maps the function $f$ with $2^n$ eigenvalues that model the $2^n$ values of $f$. The optimal solution (i.e. the extremal value of $f$) is an eigenvalue of $H$ satisfying:
$$H.|x\rangle = f(x)|x\rangle$$
Because a Hamiltonian is a Hermitian operator, it has a spectral decomposition: $H = \sum_i \lambda_i |e_i\rangle\langle e_i|$ where $|e_i\rangle$ is the $i^{th}$ basis vector and $\lambda_i$ is a real eigenvalue.

The Hamiltonian is defined with Pauli operator-basis and takes advantages of the Pauli $Z$ leading to expression
$$H = \alpha_0.Id + \alpha_1.Z_0.Z_1 + \alpha_2.Z_0.Z_2 \ldots$$
where $\alpha_i$ are real numbers.

Note that Hadfield in 2021 (Hadfield, 2021) gives a concise description of rules for composing Hamiltonians representing clauses that model clauses or functions including Boolean formulas. A Hamiltonian is implemented into a quantum circuit by deriving $U_H(t) = e^{-i.t.H}$ with $t \in [0; 2\pi]$ and using both CNOT and Z-rotations. $t$ refers to the weight in the iterative search process of QAOA.

The Trotterization defines:
$$e^{-i.t.H} = e^{-i.t.(H_1+\cdots+H_n)} \simeq \prod_{p=1}^{n} e^{-i.t.H_p}$$
with $H_p = U_1 \otimes U_2 \otimes \ldots \otimes U_n$ and $U_k \in \{Id, X, Y, Z\}$ and the difficulties in $e^{-i.t.H_p}$ lie on the computation of expression such that $e^{-i.t.U_1 \otimes U_2 \otimes \ldots \otimes U_n}$ that is the exponentiation of $H_p$.
For example, we address:
$$H_p = Z_1.Y_3.X_6.Z_8$$
$$H_p = Y_2.X_8.X_{12}.Z_{18}$$
$$H_p = Y_2.Y_3.X_8.X_{12}.Z_{18}$$

## 2. Basic concepts for $e^{-i.t.Z_j.Y_k.X_m}$ computation

2.1. Pauli gate properties

If $A^2 = Id$ then $e^{i.t.A} = \cos t . Id + i.\sin t . A$

For example, $(Z \otimes Z)^2 = Z^2 \otimes Z^2 = Id \otimes Id$ and
$$e^{i.t.(Z \otimes Z)} = \cos t . (Id \otimes Id) + i.\sin t . (Z \otimes Z)$$

Let us consider $V$ and $W$ one Pauli gate $X, Y, Z$ with $V \neq W$.
We have $R_X(-2.t) = e^{i.t.X}, R_Y(-2.t) = e^{i.t.Y}, R_Z(-2.t) = e^{i.t.Z}$ and $e^{i.t.Id} = (e^{i.t}).Id$.
So we can state that:
$$e^{i.t.A} = \cos t . Id + i.\sin t . A$$

**Definition**
A quantum circuit $Q$ is a proper exponentiation of a Hamiltonian $H$ if the matrix representation of $e^{i.\theta.H}$ is identical to the matrix representation of $Q$.

**Example**
Let us consider $H_p = X_1.Z_2$ and $e^{-i.t.X_1.Z_2}$ with 2 qubits only
Let us prove that $CZ_{1,2}.(R_X(-2.t) \otimes Id).CZ_{1,2}$ is the quantum circuit of $e^{-i.\gamma.X_1.Z_2}$

**Proof.**
We have
$$CZ_{1,2} = \begin{pmatrix} 1 & 0 & 0 & 0 \\ 0 & 1 & 0 & 0 \\ 0 & 0 & 1 & 0 \\ 0 & 0 & 0 & -1 \end{pmatrix}$$

and
$$(R_X(-2.t) \otimes Id) = \begin{pmatrix} \cos t & i.\sin t \\ i.\sin t & \cos t \end{pmatrix} \otimes \begin{pmatrix} 1 & 0 \\ 0 & 1 \end{pmatrix}$$

$$(R_X(-2.t) \otimes Id).CZ_{1,2} = \begin{pmatrix} \cos t & & i.\sin t & 0 \\ & \cos t & 0 & i.\sin t \\ i.\sin t & 0 & \cos t & \\ 0 & i.\sin t & & \cos t \end{pmatrix} . \begin{pmatrix} 1 & 0 & 0 & 0 \\ 0 & 1 & 0 & 0 \\ 0 & 0 & 1 & 0 \\ 0 & 0 & 0 & -1 \end{pmatrix}$$

$$(R_X(2.t) \otimes Id).CZ_{1,2} = \begin{pmatrix} \cos t & & i.\sin t & 0 \\ & \cos t & 0 & -i.\sin t \\ i.\sin t & 0 & \cos t & \\ 0 & i.\sin t & & -\cos t \end{pmatrix}$$

and
$$CZ_{1,2}.(R_X(2.t) \otimes Id).CZ_{1,2} = \begin{pmatrix} 1 & 0 & 0 & 0 \\ 0 & 1 & 0 & 0 \\ 0 & 0 & 1 & 0 \\ 0 & 0 & 0 & -1 \end{pmatrix} . \begin{pmatrix} \cos t & & i.\sin t & 0 \\ & \cos t & 0 & -i.\sin t \\ i.\sin t & 0 & \cos t & \\ 0 & i.\sin t & & -\cos t \end{pmatrix}$$

$$CZ_{1,2}.(R_X(2.t) \otimes Id).CZ_{1,2} = \begin{pmatrix} \cos 2 & & i.\sin t & 0 \\ & \cos t & 0 & -i.\sin t \\ i.\sin 2 & 0 & \cos t & \\ 0 & -i.\sin t & & \cos t \end{pmatrix}$$

On other hand
$$X_1.Z_2 = \begin{pmatrix} & & 1 & \\ & & & 1 \\ 1 & & & \\ & 1 & & \end{pmatrix} . \begin{pmatrix} 1 & & & \\ & -1 & & \\ & & 1 & \\ & & & -1 \end{pmatrix} = \begin{pmatrix} & & 1 & \\ & & & 1 \\ 1 & & & \\ & 1 & & \end{pmatrix}$$

and
$$e^{i.t.X_1.Z_2} = \cos t . Id + i.\sin t . (X_1.Z_2)$$

$$e^{i.t.X_1.Z_2} = \cos\gamma . \begin{pmatrix} 1 & & & \\ & 1 & & \\ & & 1 & \\ & & & 1 \end{pmatrix} + i.\sin\gamma . \begin{pmatrix} & & 1 & \\ & & & 1 \\ 1 & & & \\ & 1 & & \end{pmatrix}$$

$$e^{i.t.X_1.Z_2} = \begin{pmatrix} \cos t & 0 & i.\sin t & 0 \\ 0 & \cos t & 0 & -i.\sin t \\ i.\sin t & 0 & \cos t & 0 \\ 0 & -i.\sin t & 0 & \cos t \end{pmatrix}$$

So we conclude
$$e^{i.t.X_1.Z_2} = CZ_{1,2}.(R_X(-2.t) \otimes Id).CZ_{1,2}$$

In the literature, several common exponentiations are used including for example
$$CX_{1,2}.[Id \otimes R_Z(-2.t)].CX_{1,2} = CX_{2,1}.[R_Z(-2.t) \otimes Id].CX_{2,1} = e^{i.t.(Z \otimes Z)}$$

The calculation of $e^{-i.t.Z_{j_p}.Y_{k_p}.X_{m_p}}$ is tedious, and verifying the calculations requires manipulating matrices whose size increases rapidly with the number of qubits. Therefore, there is a need to have a more compact representation of the calculations so that the corresponding circuit can be automated. We propose to use calculations using tensor products and a simple understandable algorithm to obtain the circuit.

2.2. Proposition

**Theorem 1**

$(Id^{\otimes n-2} \otimes CX_{n,n-1}).(Id^{\otimes n-3} \otimes CX_{n-1,n-2} \otimes Id) \ldots (C_{2,1}X \otimes Id^{\otimes n-2}).R_Z(2t).(C_{2,1}X \otimes Id^{\otimes n-2}) \ldots (Id^{\otimes n-3} \otimes CX_{n-1,n-2} \otimes Id).(Id^{\otimes n-2} \otimes CX_{n,n-1}) = e^{-i.t.Z \otimes Z \otimes \ldots \otimes Z}$
$= e^{-i.t.Z^{\otimes n}}$

**Demonstration**

In the specific case of $n = 1$
Since we have $e^{-i.t.Z} = R_Z(2.t)$
We have

$$R_Z(2.t) = \begin{pmatrix} e^{-i.\frac{t}{2}} & 0 \\ 0 & e^{i.\frac{t}{2}} \end{pmatrix}$$

$$R_Z(2.t) = \begin{pmatrix} \cos\frac{t}{2} & 0 \\ 0 & \cos\frac{t}{2} \end{pmatrix} - i.\begin{pmatrix} \sin\frac{t}{2} & 0 \\ 0 & -\sin\frac{t}{2} \end{pmatrix}$$

$$R_Z(2.t).|0\rangle = \cos t . Id - i.\sin t . Z = e^{-i.t.Z}$$

Let us assume that

$(Id^{\otimes n-2} \otimes CX_{n,n-1}).(Id^{\otimes n-3} \otimes CX_{n-1,n-2} \otimes Id) \ldots (C_{2,1}X \otimes Id^{\otimes n-2}).R_Z(2t).(C_{2,1}X \otimes Id^{\otimes n-2}) \ldots (Id^{\otimes n-3} \otimes CX_{n-1,n-2} \otimes Id).(Id^{\otimes n-2} \otimes CX_{n,n-1}) = e^{-i.t.Z \otimes Z \otimes \ldots \otimes Z}$
$= e^{-i.t.Z^{\otimes n}}$

is true and let us demonstrate that the equality is true for $n + 1$.

By hypothesis the following equality is assume to be valid:

$(Id^{\otimes n-2} \otimes CX_{n,n-1}).(Id^{\otimes n-3} \otimes CX_{n-1,n-2} \otimes Id) \ldots (CX_{2,1} \otimes Id^{\otimes n-2}).[R_Z(2t) \otimes Id^{\otimes n-1}].(CX_{2,1} \otimes Id^{\otimes n-2}) \ldots (Id^{\otimes n-3} \otimes CX_{n-1,n-2} \otimes Id).(Id^{\otimes n-2} \otimes CX_{n,n-1}) = e^{-i.t.Z^{\otimes n}}$

By consequence we can multiply by $(Id^{\otimes n-1} \otimes CX_{n+1,n})$ at the left and right of the equality:

$(Id^{\otimes n-1} \otimes CX_{n+1,n}).[(Id^{\otimes n-2} \otimes CX_{n,n-1} \otimes Id).(Id^{\otimes n-3} \otimes CX_{n-1,n-2} \otimes Id \otimes Id) \ldots (CX_{2,1} \otimes Id^{\otimes n-1}).[R_Z(2t) \otimes Id^{\otimes n}].(CX_{2,1} \otimes Id^{\otimes n-1}) \ldots (Id^{\otimes n-3} \otimes CX_{n-1,n-2} \otimes Id \otimes Id).(Id^{\otimes n-2} \otimes CX_{n,n-1} \otimes Id)].(Id^{\otimes n-1} \otimes CX_{n+1,n})$

which can be rewritten:

$$H_1 = (Id^{\otimes n-1} \otimes CX_{n+1,n}).\left[e^{-i.t.Z^{\otimes n}} \otimes Id\right].(Id^{\otimes n-1} \otimes CX_{n+1,n})$$

It is possible to apply this circuit to $|q_1 q_2 \ldots q_{n-1} q_n q_{n+1}\rangle$ considering :
- first $|q_1 q_2 \ldots q_{n-1} q_n 0\rangle$
- second $|q_1 q_2 \ldots q_{n-1} 01\rangle$
- last $|q_1 q_2 \ldots q_{n-1} 11\rangle$

**Let us consider $|q_1 q_2 \ldots q_{n-1} q_n 0\rangle$**

We have

$H_1.|q_1 q_2 \ldots q_{n-1} q_n 0\rangle = (Id^{\otimes n-1} \otimes CX_{n+1,n}).\left[e^{-i.t.Z^{\otimes n}} \otimes Id\right].(Id^{\otimes n-1} \otimes CX_{n+1,n}).|q_1 q_2 \ldots q_n 0\rangle$

$H_1.|q_1 q_2 \ldots q_{n-1} q_n 0\rangle = (Id^{\otimes n-1} \otimes CX_{n+1,n}).\left[e^{-i.t.Z^{\otimes n}} \otimes Id\right].|q_1 q_2 \ldots q_n 0\rangle$

$H_1.|q_1 q_2 \ldots q_{n-1} q_n 0\rangle$
$= (Id^{\otimes n-1} \otimes CX_{n+1,n}).[\cos(t).|q_1 q_2 \ldots q_n\rangle \otimes |0\rangle - i.\sin(t).(Z^{\otimes n}.|q_1 q_2 \ldots q_n\rangle) \otimes |0\rangle]$

Because $Z.|0\rangle = |0\rangle$, we have
$$H_1.|q_1q_2...q_{n-1}q_n0\rangle = \cos(t).|q_1q_2...q_n\rangle \otimes |0\rangle - i.\sin(t).(Z^{\otimes n}.|q_1q_2...q_n\rangle) \otimes Z.|0\rangle$$

$$H_1.|q_1q_2...q_{n-1}q_n0\rangle = \cos(t).|q_1q_2...q_n0\rangle - i.t.\sin(t).(Z^{\otimes n+1}.|q_1q_2...q_n0\rangle)$$
$$H_1.|q_1q_2...q_{n-1}q_n0\rangle = e^{-i.t.Z^{\otimes n+1}}.|q_1q_2...q_n0\rangle$$

This result prove that $H_1$ applied to any quantum state where the last qubit in position $n+1$ is 0 compute $e^{-i.t.Z^{\otimes n+1}}$.

**Let us consider $|q_1q_2...q_{n-1}01\rangle$**
We have
$$H_1.|q_1q_2...q_{n-1}01\rangle = (Id^{\otimes n-1} \otimes CX_{n+1,n}).[e^{-i.t.Z^{\otimes n}} \otimes Id].(Id^{\otimes n-1} \otimes CX_{n+1,n}).|q_1q_2...q_{n-1}01\rangle$$

Because $X.|0\rangle = |1\rangle$ we have:
$$H_1.|q_1q_2...q_{n-1}01\rangle = (Id^{\otimes n-1} \otimes CX_{n+1,n}).[e^{-i.t.Z^{\otimes n}} \otimes Id].|q_1q_2...q_{n-1}11\rangle$$

so
$$H_1.|q_1q_2...q_{n-1}01\rangle$$
$$= (Id^{\otimes n-1} \otimes CX_{n+1,n}).[\cos(t).|q_1q_2...q_{n-1}1\rangle \otimes |1\rangle - i.\sin(t).(Z^{\otimes n}.|q_1q_2...q_{n-1}1\rangle)$$
$$\otimes |1\rangle]$$

Because $Z.|0\rangle = |0\rangle$ and $Z.|1\rangle = -|1\rangle$, we have :
$$H_1.|q_1q_2...q_{n-1}01\rangle$$
$$= (Id^{\otimes n-1} \otimes CX_{n+1,n}).[\cos(t).|q_1q_2...q_{n-1}\rangle + i.\sin(t).(Z^{\otimes n-1}.|q_1q_2...q_{n-1}\rangle)] \otimes |11\rangle$$

and
$$H_1.|q_1q_2...q_{n-1}01\rangle = \cos(t).|q_1q_2...q_{n-1}01\rangle - i.\sin(t).(Z^{\otimes n+1})|q_1q_2...q_{n-1}01\rangle$$
$$H_1.|q_1q_2...q_{n-1}01\rangle = e^{-i.t.Z^{\otimes n+1}}.|q_1q_2...q_n01\rangle$$

This result prove that $H_1$ applied to any quantum state where the two last qubits is $|01\rangle$ computes $e^{-i.t.Z^{\otimes n+1}}$.

**Let us consider $|q_1q_2...q_{n-1}11\rangle$**
We have
$$H_1.|q_1q_2...q_{n-1}11\rangle = (Id^{\otimes n-1} \otimes CX_{n+1,n}).[e^{-i.t.Z^{\otimes n}} \otimes Id].(Id^{\otimes n-1} \otimes CX_{n+1,n}).|q_1q_2...q_{n-1}11\rangle$$
$$H_1.|q_1q_2...q_{n-1}11\rangle = (Id^{\otimes n-1} \otimes CX_{n+1,n}).[e^{-i.t.Z^{\otimes n}} \otimes Id].|q_1q_2...q_{n-1}01\rangle$$

and
$$H_1.|q_1q_2...q_{n-1}11\rangle$$
$$= (Id^{\otimes n-1} \otimes CX_{n+1,n}).[\cos(t).|q_1q_2...q_{n-1}0\rangle \otimes |1\rangle - i.\sin(t).(Z^{\otimes n}.|q_1q_2...q_{n-1}0\rangle)$$
$$\otimes |1\rangle]$$

Because $Z.|0\rangle = |0\rangle$ we have:
$$H_1.|q_1q_2...q_{n-1}11\rangle$$
$$= (Id^{\otimes n-1} \otimes CX_{n+1,n}).[\cos(t).|q_1q_2...q_{n-1}\rangle - i.\sin(t).(Z^{\otimes n-1}.|q_1q_2...q_{n-1}\rangle)] \otimes |01\rangle$$
$$H_1.|q_1q_2...q_{n-1}11\rangle = [\cos(t).|q_1q_2...q_{n-1}\rangle - i.\sin(t).(Z^{\otimes n-1}.|q_1q_2...q_{n-1}\rangle)] \otimes |11\rangle$$

Since $Z.|1\rangle = -|1\rangle$ we have $Z \otimes Z.|11\rangle = |11\rangle$
and
$$H_1.|q_1q_2...q_{n-1}11\rangle = \cos(t).|q_1q_2...q_{n-1}11\rangle - i.t.\sin(t).(Z^{\otimes n+1})$$
$$H_1.|q_1q_2...q_{n-1}11\rangle = e^{-i.t.Z^{\otimes n+1}}.|q_1q_2...q_n01\rangle$$

This result proves that $H_1$ applied to any quantum state where the two last qubits is $|11\rangle$ computes $e^{-i.t.Z^{\otimes n+1}}$.
□

**Remark**
A quasi-direct application gives, for example:
$$Q = e^{-i.t.Z_j.Z_k.Z_m} = CX_{jk}.CX_{km}.RZ_m(2.t).CX_{km}.CX_{jk}$$

with
$$Q = e^{-i.t.Z_j.Z_k.Z_m} = e^{-i.t(...\otimes...\otimes Z\otimes...\otimes Z\otimes...\otimes Z\otimes...)}$$
where the Pauli gate $Z$ is applied on qubits number $j,k,m$

□

So it is possible to define the quantum circuit for $e^{-i.t.Z^{\otimes n}}$.

**Theorem 2**
Any $e^{-i.t.U_1\otimes U_2\otimes...\otimes U_n}$ where $U_k = X$ can be defined by application of $(Id^{\otimes k-1} \otimes H \otimes Id^{\otimes n-k})$ before and after $e^{-i.t.U_1\otimes U_2\otimes..U_{k-1}\otimes Z\otimes U_{k+1}...\otimes U_n}$.

**Demonstration**
Let us consider
$$H = (Id^{\otimes k-1} \otimes H \otimes Id^{\otimes n-k}).e^{-i.t.U_1\otimes U_2\otimes...\otimes U_n}.(Id^{\otimes k-1} \otimes H \otimes Id^{\otimes n-k})$$
Where the operator $H$ is applied at the $k^{th}$ qubit.
Two situations can hold depending on $|q_k\rangle$.

**Case 1.** $|q_k\rangle = 0$
So we have:
$$H.|q_1, q_2, ..., q_{k-1}, 0, q_{k+1}, ..., q_n\rangle = (Id^{\otimes k-1} \otimes H \otimes Id^{\otimes n-k}).e^{-i.t.U_1\otimes U_2\otimes...\otimes U_n}.(Id^{\otimes k-1} \otimes H \otimes Id^{\otimes n-k})|q_1, q_2, ..., q_{k-1}, 0, q_{k+1}, ..., q_n\rangle$$

Because $H.|0\rangle = |p\rangle = \frac{|0\rangle+|1\rangle}{\sqrt{2}}$:
$$H.|q_1, q_2, ..., q_{k-1}, 0, q_{k+1}, ..., q_n\rangle = (Id^{\otimes k-1} \otimes H \otimes Id^{\otimes n-k}).e^{-i.t.U_1\otimes U_2\otimes...\otimes U_n}.|q_1, q_2, ..., q_{k-1}, p, q_{k+1}, ..., q_n\rangle$$

and
$$H.|q_1, q_2, ..., q_{k-1}, 0, q_{k+1}, ..., q_n\rangle = (Id^{\otimes k-1} \otimes H \otimes Id^{\otimes n-k}).[\cos(t).Id^{\otimes n} - i.\sin(t).(U_1 \otimes U_2 \otimes ... \otimes U_n)].|q_1, q_2, ..., q_{k-1}, p, q_{k+1}, ..., q_n\rangle$$

Because we assume $U_k = Z$ we have:
$$H.|q_1, q_2, ..., q_{k-1}, 0, q_{k+1}, ..., q_n\rangle = (Id^{\otimes k-1} \otimes H \otimes Id^{\otimes n-k}).[\cos(t).|q_1, q_2, ..., q_{k-1}, p, q_{k+1}, ..., q_n\rangle - i.\sin(t).|U_1.q_1, U_2.q_2, ..., U_{k-1}.q_{k-1}, Z.p, U_{k+1}.q_{k+1}, ..., U_n.q_n\rangle]$$

Because $Z.|p\rangle = |m\rangle = \frac{|0\rangle-|1\rangle}{\sqrt{2}}$
$$H.|q_1, q_2, ..., q_{k-1}, 0, q_{k+1}, ..., q_n\rangle = (Id^{\otimes k-1} \otimes H \otimes Id^{\otimes n-k}).[\cos(t).|q_1, q_2, ..., q_{k-1}, p, q_{k+1}, ..., q_n\rangle - i.\sin(t).|U_1.q_1, U_2.q_2, ..., U_{k-1}.q_{k-1}, m, U_{k+1}.q_{k+1}, ..., U_n.q_n\rangle]$$

Because $H|m\rangle = |1\rangle$ and $H|p\rangle = |0\rangle$
$$H.|q_1, q_2, ..., q_{k-1}, 0, q_{k+1}, ..., q_n\rangle = \cos(t).|q_1, q_2, ..., q_{k-1}, 0, q_{k+1}, ..., q_n\rangle - i.\sin(t).|U_1.q_1, U_2.q_2, ..., U_{k-1}.q_{k-1}, 1, U_{k+1}.q_{k+1}, ..., U_n.q_n\rangle$$

Because $X|0\rangle = |1\rangle$ we have:
$$H.|q_1, q_2, ..., q_{k-1}, 0, q_{k+1}, ..., q_n\rangle = \cos(t).|q_1, q_2, ..., q_{k-1}, 0, q_{k+1}, ..., q_n\rangle - i.\sin(t).|U_1.q_1, U_2.q_2, ..., U_{k-1}.q_{k-1}, X.0, U_{k+1}.q_{k+1}, ..., U_n.q_n\rangle$$

and
$$H.|q_1, q_2, ..., q_{k-1}, 0, q_{k+1}, ..., q_n\rangle = e^{-i.t.V_1\otimes V_2\otimes...\otimes V_n}.|q_1, q_2, ..., q_{k-1}, 0, q_{k+1}, ..., q_n\rangle$$
with $V_j = U_j$ if $j \neq k$ and $V_k = X$

**Case 2.** $|q_k\rangle = 1$
So we have:

$$H.|q_1, q_2, \ldots, q_{k-1}, 1, q_{k+1}, \ldots, q_n\rangle$$
$$= (Id^{\otimes k-1} \otimes H \otimes Id^{\otimes n-k}). e^{-i.t.U_1 \otimes U_2 \otimes \ldots \otimes U_n}. (Id^{\otimes k-1} \otimes H \otimes Id^{\otimes n-k})|q_1, q_2, \ldots, q_{k-1}, 1, q_{k+1}, \ldots, q_n\rangle$$

Because $H.|1\rangle = |m\rangle$ we have:
$$H.|q_1, q_2, \ldots, q_{k-1}, 1, q_{k+1}, \ldots, q_n\rangle$$
$$= (Id^{\otimes k-1} \otimes H \otimes Id^{\otimes n-k}). e^{-i.t.U_1 \otimes U_2 \otimes \ldots \otimes U_n}. |q_1, q_2, \ldots, q_{k-1}, m, q_{k+1}, \ldots, q_n\rangle$$

then
$$H.|q_1, q_2, \ldots, q_{k-1}, 1, q_{k+1}, \ldots, q_n\rangle$$
$$= (Id^{\otimes k-1} \otimes H \otimes Id^{\otimes n-k}). [\cos(t). Id^{\otimes n} - i.\sin(t). (U_1 \otimes U_2 \otimes \ldots \otimes U_n)]. |q_1, q_2, \ldots, q_{k-1}, m, q_{k+1}, \ldots, q_n\rangle$$

and
$$H.|q_1, q_2, \ldots, q_{k-1}, 1, q_{k+1}, \ldots, q_n\rangle$$
$$= (Id^{\otimes k-1} \otimes H \otimes Id^{\otimes n-k}). [\cos(t). |q_1, q_2, \ldots, q_{k-1}, m, q_{k+1}, \ldots, q_n\rangle - i.\sin(t). |U_1.q_1, U_2.q_2, \ldots, U_{k-1}.q_{k-1}, Z.m, U_{k+1}.q_{k+1}, \ldots, U_n.q_n\rangle]$$

Because $Z.|m\rangle = |p\rangle$
$$H.|q_1, q_2, \ldots, q_{k-1}, 1, q_{k+1}, \ldots, q_n\rangle$$
$$= (Id^{\otimes k-1} \otimes H \otimes Id^{\otimes n-k}). [\cos(t). |q_1, q_2, \ldots, q_{k-1}, m, q_{k+1}, \ldots, q_n\rangle - i.\sin(t). |U_1.q_1, U_2.q_2, \ldots, U_{k-1}.q_{k-1}, p, U_{k+1}.q_{k+1}, \ldots, U_n.q_n\rangle]$$

Because $H|p\rangle = |0\rangle$ and $H|m\rangle = |1\rangle$, we have:
$$H.|q_1, q_2, \ldots, q_{k-1}, 1, q_{k+1}, \ldots, q_n\rangle$$
$$= \cos(t). |q_1, q_2, \ldots, q_{k-1}, 1, q_{k+1}, \ldots, q_n\rangle - i.\sin(t). |U_1.q_1, U_2.q_2, \ldots, U_{k-1}.q_{k-1}, 0, U_{k+1}.q_{k+1}, \ldots, U_n.q_n\rangle$$

So because $X.|1\rangle = |0\rangle$:
$$H.|q_1, q_2, \ldots, q_{k-1}, 1, q_{k+1}, \ldots, q_n\rangle$$
$$= \cos(t). |q_1, q_2, \ldots, q_{k-1}, 1, q_{k+1}, \ldots, q_n\rangle - i.\sin(t). |U_1.q_1, U_2.q_2, \ldots, U_{k-1}.q_{k-1}, X.1, U_{k+1}.q_{k+1}, \ldots, U_n.q_n\rangle$$

and
$$H.|q_1, q_2, \ldots, q_{k-1}, 1, q_{k+1}, \ldots, q_n\rangle$$
$$= \cos(t). |q_1, q_2, \ldots, q_{k-1}, 1, q_{k+1}, \ldots, q_n\rangle - i.\sin(t). |U_1.q_1, U_2.q_2, \ldots, U_{k-1}.q_{k-1}, X.1, U_{k+1}.q_{k+1}, \ldots, U_n.q_n\rangle$$

and
$$H.|q_1, q_2, \ldots, q_{k-1}, 1, q_{k+1}, \ldots, q_n\rangle = e^{-i.t.V_1 \otimes V_2 \otimes \ldots \otimes V_n}. |q_1, q_2, \ldots, q_{k-1}, 1, q_{k+1}, \ldots, q_n\rangle$$
with $V_j = U_j$ if $j \neq k$ and $V_k = X$.

□

**Theorem 3**

Any $e^{-i.t.U_1 \otimes U_2 \otimes \ldots \otimes U_n}$ where $U_k = Y$ can defined by application of $(Id^{\otimes k-1} \otimes S \otimes Id^{\otimes n-k})$ before and $(Id^{\otimes k-1} \otimes S^\dagger \otimes Id^{\otimes n-k})$ after $e^{-i.t.U_1 \otimes U_2 \otimes ..U_{k-1} \otimes X \otimes U_{k+1} \ldots \otimes U_n}$.

**Demonstration**
Let us consider
$$H = (Id^{\otimes k-1} \otimes S \otimes Id^{\otimes n-k}). e^{-i.t.U_1 \otimes U_2 \otimes \ldots \otimes U_n}. (Id^{\otimes k-1} \otimes S^\dagger \otimes Id^{\otimes n-k})$$
Two situations can hold depending on $|q_k\rangle$.

**Remark**

$$S.|0\rangle = |0\rangle, S.|1\rangle = i.|1\rangle$$
$$S^\dagger.|0\rangle = |0\rangle, S^\dagger.|1\rangle = -i.|1\rangle$$
$$Y.|0\rangle = i.|1\rangle, Y.|1\rangle = -i.|0\rangle \text{ and } |1\rangle = -i.Y.|0\rangle, |0\rangle = i.Y.|1\rangle$$

**Case 1.** $|q_k\rangle = 0$

$$H.|q_1, q_2, \dots, q_{k-1}, 0, q_{k+1}, \dots, q_n\rangle$$
$$= (Id^{\otimes k-1} \otimes S \otimes Id^{\otimes n-k}).e^{-i.t.U_1 \otimes U_2 \otimes \dots \otimes U_n}.(Id^{\otimes k-1} \otimes S^\dagger \otimes Id^{\otimes n-k})$$
$$.|q_1, q_2, \dots, q_{k-1}, 0, q_{k+1}, \dots, q_n\rangle$$

Because $S^\dagger.|0\rangle = |0\rangle$, we have:
$$H.|q_1, q_2, \dots, q_{k-1}, 0, q_{k+1}, \dots, q_n\rangle$$
$$= (Id^{\otimes k-1} \otimes S \otimes Id^{\otimes n-k}).e^{-i.t.U_1 \otimes U_2 \otimes \dots \otimes U_n}.|q_1, q_2, \dots, q_{k-1}, 0, q_{k+1}, \dots, q_n\rangle$$

and
$$H.|q_1, q_2, \dots, q_{k-1}, 0, q_{k+1}, \dots, q_n\rangle$$
$$= (Id^{\otimes k-1} \otimes S \otimes Id^{\otimes n-k}).[\cos(t).Id^{\otimes n}$$
$$- i.\sin(t).(U_1 \otimes U_2 \otimes \dots \otimes U_n)].|q_1, q_2, \dots, q_{k-1}, 0, q_{k+1}, \dots, q_n\rangle$$

So
$$H.|q_1, q_2, \dots, q_{k-1}, 0, q_{k+1}, \dots, q_n\rangle$$
$$= (Id^{\otimes k-1} \otimes S \otimes Id^{\otimes n-k}).[\cos(t).|q_1, q_2, \dots, q_{k-1}, 0, q_{k+1}, \dots, q_n\rangle$$
$$- i.\sin(t).|U_1.q_1, U_2.q_2, \dots, U_{k-1}.q_{k-1}, X.0, U_{k+1}.q_{k+1}, \dots, U_n.q_n\rangle]$$

Then because $X.|0\rangle = |1\rangle$
$$H.|q_1, q_2, \dots, q_{k-1}, 0, q_{k+1}, \dots, q_n\rangle$$
$$= (Id^{\otimes k-1} \otimes S \otimes Id^{\otimes n-k}).[\cos(t).|q_1, q_2, \dots, q_{k-1}, 0, q_{k+1}, \dots, q_n\rangle$$
$$- i.\sin(t).|U_1.q_1, U_2.q_2, \dots, U_{k-1}.q_{k-1}, 1, U_{k+1}.q_{k+1}, \dots, U_n.q_n\rangle]$$

Because $S|1\rangle = i.|1\rangle$ and $S|0\rangle = |0\rangle$
$$H.|q_1, q_2, \dots, q_{k-1}, 0, q_{k+1}, \dots, q_n\rangle$$
$$= \cos(t).|q_1, q_2, \dots, q_{k-1}, 0, q_{k+1}, \dots, q_n\rangle$$
$$+ \sin(t).|U_1.q_1, U_2.q_2, \dots, U_{k-1}.q_{k-1}, 1, U_{k+1}.q_{k+1}, \dots, U_n.q_n\rangle$$

then because $-i.Y.|0\rangle = |1\rangle$, we have :
$$H.|q_1, q_2, \dots, q_{k-1}, 0, q_{k+1}, \dots, q_n\rangle$$
$$= \cos(t).|q_1, q_2, \dots, q_{k-1}, 0, q_{k+1}, \dots, q_n\rangle$$
$$- i.\sin(t).|U_1.q_1, U_2.q_2, \dots, U_{k-1}.q_{k-1}, Y.0, U_{k+1}.q_{k+1}, \dots, U_n.q_n\rangle$$

and
$$H.|q_1, q_2, \dots, q_{k-1}, 0, q_{k+1}, \dots, q_n\rangle = e^{-i.t.V_1 \otimes V_2 \otimes \dots \otimes V_n}.|q_1, q_2, \dots, q_{k-1}, 0, q_{k+1}, \dots, q_n\rangle$$
with $V_j = U_j$ if $j \neq k$ and $V_k = Y$.

**Case 2.** $|q_k\rangle = 1$
We have:
$$H.|q_1, q_2, \dots, q_{k-1}, 1, q_{k+1}, \dots, q_n\rangle$$
$$= (Id^{\otimes k-1} \otimes S \otimes Id^{\otimes n-k}).e^{-i.t.U_1 \otimes U_2 \otimes \dots \otimes U_n}.(Id^{\otimes k-1} \otimes S^\dagger \otimes Id^{\otimes n-k})$$
$$.|q_1, q_2, \dots, q_{k-1}, 1, q_{k+1}, \dots, q_n\rangle$$

For $S^\dagger.|1\rangle = -i.|1\rangle$:
$$H.|q_1, q_2, \dots, q_{k-1}, 1, q_{k+1}, \dots, q_n\rangle$$
$$= -i.(Id^{\otimes k-1} \otimes S \otimes Id^{\otimes n-k}).e^{-i.t.U_1 \otimes U_2 \otimes \dots \otimes U_n}.|q_1, q_2, \dots, q_{k-1}, 1, q_{k+1}, \dots, q_n\rangle$$

then
$$H.|q_1, q_2, \dots, q_{k-1}, 1, q_{k+1}, \dots, q_n\rangle$$
$$= -i.(Id^{\otimes k-1} \otimes S \otimes Id^{\otimes n-k}).[\cos(t).Id^{\otimes n}$$
$$- i.\sin(t).(U_1 \otimes U_2 \otimes \dots \otimes U_n)].|q_1, q_2, \dots, q_{k-1}, 1, q_{k+1}, \dots, q_n\rangle$$

and
$$H.|q_1, q_2, \dots, q_{k-1}, 1, q_{k+1}, \dots, q_n\rangle$$
$$= -i.(Id^{\otimes k-1} \otimes S \otimes Id^{\otimes n-k}).[\cos(t).|q_1, q_2, \dots, q_{k-1}, 1, q_{k+1}, \dots, q_n\rangle$$
$$- i.\sin(t).|U_1.q_1, U_2.q_2, \dots, U_{k-1}.q_{k-1}, X.1, U_{k+1}.q_{k+1}, \dots, U_n.q_n\rangle]$$

Because $X.|1\rangle = |0\rangle$ we have:
$$H.|q_1, q_2, \dots, q_{k-1}, 1, q_{k+1}, \dots, q_n\rangle$$
$$= (Id^{\otimes k-1} \otimes S \otimes Id^{\otimes n-k}).[-i.\cos(t).|q_1, q_2, \dots, q_{k-1}, 1, q_{k+1}, \dots, q_n\rangle$$
$$- \sin(t).|U_1.q_1, U_2.q_2, \dots, U_{k-1}.q_{k-1}, 0, U_{k+1}.q_{k+1}, \dots, U_n.q_n\rangle]$$

Since we have $S|1\rangle = i.|1\rangle$ and $S|0\rangle = |0\rangle$

$$H.|q_1, q_2, \ldots, q_{k-1}, 1, q_{k+1}, \ldots, q_n\rangle$$
$$= \cos(t).|q_1, q_2, \ldots, q_{k-1}, 1, q_{k+1}, \ldots, q_n\rangle$$
$$- \sin(t).|U_1.q_1, U_2.q_2, \ldots, U_{k-1}.q_{k-1}, 0, U_{k+1}.q_{k+1}, \ldots, U_n.q_n\rangle$$

For $Y.|1\rangle = -i.|0\rangle$ i.e. $i.Y.|1\rangle = |0\rangle$ we have
$$H.|q_1, q_2, \ldots, q_{k-1}, 1, q_{k+1}, \ldots, q_n\rangle$$
$$= \cos(t).|q_1, q_2, \ldots, q_{k-1}, 1, q_{k+1}, \ldots, q_n\rangle$$
$$- i.\sin(t).|U_1.q_1, U_2.q_2, \ldots, U_{k-1}.q_{k-1}, Y.1, U_{k+1}.q_{k+1}, \ldots, U_n.q_n\rangle$$

Then
$$H.|q_1, q_2, \ldots, q_{k-1}, 1, q_{k+1}, \ldots, q_n\rangle = e^{-i.t.V_1 \otimes V_2 \otimes \ldots \otimes V_n}.|q_1, q_2, \ldots, q_{k-1}, 1, q_{k+1}, \ldots, q_n\rangle$$
with $V_j = U_j$ for $j \neq k$ and $V_k = Y$.

□

## 2.2. Algorithm

The algorithm 1 is composed of step 1 that is the initialization and one loop starting at step 2 and finishing at step 10. Step 1 is the initialization of the quantum circuit taking into account the $Z$ gate only. The loop iterates on $H_p$ to scan sequentially the Pauli gates. If in position $k$ we have a $X$ gate, i.e. $H_p[k] = X$, then $(Id^{\otimes k-1} \otimes H \otimes Id^{\otimes n-k})$ must be added in front of the quantum circuit $Q$ and $(Id^{\otimes k-1} \otimes H \otimes Id^{\otimes n-k})$ has to be added at the end of $Q$. This is achieved at step 4:
$$Q = (Id^{\otimes k-1} \otimes H \otimes Id^{\otimes n-k}).Q.(Id^{\otimes k-1} \otimes H \otimes Id^{\otimes n-k})$$
This update inserts $H$ at qubit $k$ at rear and front of $Q$ to define: $Q = H_k.Q.H_k$.

If in position $k$ we have a $Y$ gate to update are achieved. The first one at step 7 which is the same operation than operation achieved at step 4 when a $X$ gate is processed:
$$Q = (Id^{\otimes k-1} \otimes H \otimes Id^{\otimes n-k}).Q.(Id^{\otimes k-1} \otimes H \otimes Id^{\otimes n-k})$$
The second one at step 8 consist in adding $S$ gate in rear of $Q$ and $S^\dagger$ in front of $Q$:
$$S_k.Q.S_k^\dagger$$

---

**Algorithm 1.** *Exponentiation_of_Hamiltonian*

**Input:**
    $H_p$ : a Hamiltonian that is a tensor product of Pauli operators $X, Y, Z, Id$
    $t$ : a real parameter

**Output:**
    $Q$ : the quantum circuit of $e^{-i.t.H_p}$

**Begin**
1  Define $Q = e^{-i.t.U_1 \otimes U_2 \otimes \ldots \otimes U_n}$ with $U_j = Z$ if $H_p[j] \neq Id$ (theorem 1)
2.  **For** $k$ in $1 \ldots n$ **do**
3    **If** $(H_p[k] = X)$ **Then**
4      Let $Q = (Id^{\otimes k-1} \otimes H \otimes Id^{\otimes n-k}).Q.(Id^{\otimes k-1} \otimes H \otimes Id^{\otimes n-k}) = H_k.Q.H_k$
5.    **End If**
6    **If** $(H_p[k] = Y)$ **Then**
7      Let $Q = (Id^{\otimes k-1} \otimes H \otimes Id^{\otimes n-k}).Q.(Id^{\otimes k-1} \otimes H \otimes Id^{\otimes n-k}) = H_k.Q.H_k$
8      Let $Q = (Id^{\otimes k-1} \otimes S \otimes Id^{\otimes n-k}).Q.(Id^{\otimes k-1} \otimes S^\dagger \otimes Id^{\otimes n-k}) = S_k.Q.S_k^\dagger$
9    **End If**
10 **End For**
**End**

---

## 2.3. Example and validation with Qiskit result

Let us consider with 6 qubits only:
$$H_p = Y_2.Y_4.X_6$$
$$H_p = Id \otimes Y \otimes Id \otimes Y \otimes Id \otimes X$$

The Algorithm 1 initialize at line 1, the quantum circuit:
$$Q = e^{-i.t.(Id \otimes Z \otimes Id \otimes Z \otimes Id \otimes Z)} = CX_{6,4}.CX_{4,2}.RZ_2(2t).CX_{4,2}.CX_{6,4}$$

For $k = 2$, at line 7, the quantum circuit is updated:
$$Q = (Id \otimes H \otimes Id^{\otimes 4}).(CX_{6,4}.CX_{4,2}.RZ_2(2t).CX_{4,2}.CX_{6,4}).(Id \otimes H \otimes Id^{\otimes 4})$$
And at line 8, the circuit is:
$$Q = (Id \otimes S \otimes Id^{\otimes 4}).(Id \otimes H \otimes Id^{\otimes 4}).(CX_{6,4}.CX_{4,2}.RZ_2(2t).CX_{4,2}.CX_{6,4}).(Id \otimes H \otimes Id^{\otimes 4}).(Id \otimes S^\dagger \otimes Id^{\otimes 4})$$
$$Q = (Id \otimes S.H \otimes Id^{\otimes 4}).(CX_{6,4}.CX_{4,2}.RZ_2(2t).CX_{4,2}.CX_{6,4}).(Id \otimes H.S^\dagger \otimes Id^{\otimes 4})$$

For $k = 4$, at line 7 and 8, the quantum circuit is updated:
$$Q = (Id^{\otimes 3} \otimes S.H \otimes Id^{\otimes 2}).(Id \otimes S.H \otimes Id^{\otimes 4}).(CX_{6,4}.CX_{4,2}.RZ_2(2t).CX_{4,2}.CX_{6,4}).(Id \otimes H.S^\dagger \otimes Id^{\otimes 4}).(Id^{\otimes 3} \otimes H.S^\dagger \otimes Id^{\otimes 2})$$
That is
$$Q = (S_2.H_2.S_4.H_4).(CX_{6,4}.CX_{4,2}.RZ_2(2t).CX_{4,2}.CX_{6,4}).(H_2.S_2^\dagger.H_4.S_4^\dagger)$$
For $k = 4$, at line 4, the quantum circuit is updated:
$$Q = (S_2.H_2.S_4.H_4.H_6).(CX_{6,4}.CX_{4,2}.RZ_2(2t).CX_{4,2}.CX_{6,4}).(H_2.S_2^\dagger.H_4.S_4^\dagger.H_6)$$

It is possible to compare the quantum circuit we obtained with the quantum circuit provided by Qiskit using the following Python code:

```python
hamiltonian3 = (1/4*X^I^Y^I^Y^I)
evo_time = Parameter('t')
evolution_op = (evo_time*hamiltonian3).exp_i()
num_time_slices = 1
trotterized_op = PauliTrotterEvolution(
                 trotter_mode='trotter',
                 reps=num_time_slices).convert(evolution_op)
trot_op_circ = trotterized_op.to_circuit()
print(trot_op_circ)
trot_op_circ_decomp = trot_op_circ.decompose()
print(trot_op_circ_decomp)
```

Note that the Hamiltonian has to be written in the reverse order of the mathematical formulation and Qiskit provides the detail of the circuit using the **decompose()** method. The result provided in Figure 1, gives a quantum circuit that meets the quantum circuit generated by application of algorithm 1.

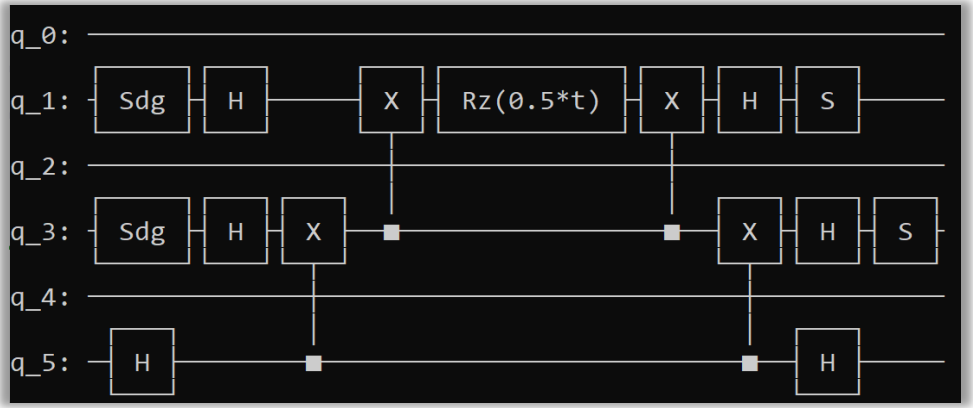

Figure 1. Qiskit exponentiation of $H_p = Y_2.Y_4.X_6$

**Remark**
A similar algorithm could be defined starting with
$$Q = e^{-i.t.U_1 \otimes U_2 \otimes ... \otimes U_n} \text{ with } U_j = X \text{ if } H_p[j] \neq Id)$$
and switching $X$ to $Y$ using operator $S$ and $S^\dagger$ and switching from $X$ to $Z$ using $CX$

**Example:**

Let us consider $H_p = X \otimes Z \otimes Z \otimes Id \otimes Y$

Step 1.
The Algorithm initialize the quantum circuit using $X$ gates only:
$$Q = e^{-i.t.(X \otimes X \otimes X \otimes Id \otimes X)}$$
that is introduced in Figure 2.

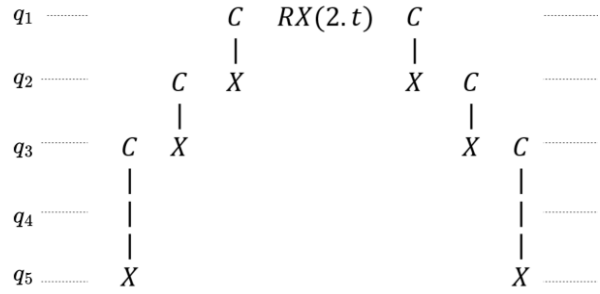

Figure 2. Quantum circuit for $e^{-i.t.(X \otimes X \otimes X \otimes Id \otimes X)}$

Step 2 Using $H$ we switch $X$ into $Z$ to generate $e^{i.t.X \otimes Z \otimes Z \otimes Id \otimes X}$ and to obtain the quantum circuit of Figure 3.

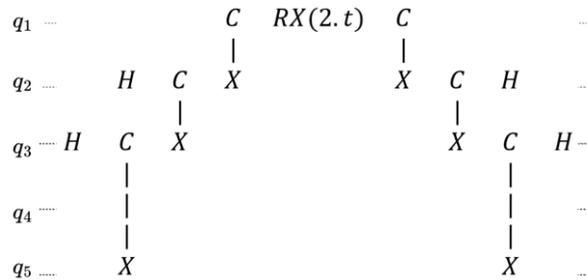

Figure 3. Quantum circuit for $e^{i.t.X \otimes Z \otimes Z \otimes Id \otimes X}$

Step 3 Using $S$ and $S^\dagger$ we switch $X$ into $Y$ to generate $e^{i.t.X \otimes Z \otimes Z \otimes Id \otimes Y}$ and to obtain the quantum circuit of Figure 4.

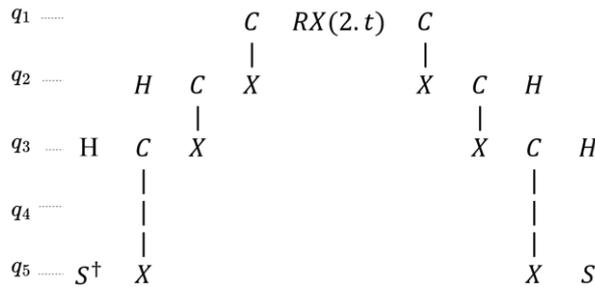

Figure 4. Quantum circuit for $e^{i.t.X \otimes Z \otimes Z \otimes Id \otimes Y}$

**Remark**
A more compact algorithm could be defined starting with
$Q = e^{-i.t.U_1 \otimes U_2 \otimes \ldots \otimes U_n}$ with $U_j = X$ if $H_p[j] \in \{X, Y\}$ and $U_j = Z$ if $H_p[j] = Z$
and switching $X$ to $Y$ using operator $S$ and $S^\dagger$

**Example:**
Let us consider $H_p = X \otimes Z \otimes Z \otimes Id \otimes Y$

Step 1.

The Algorithm initialize the quantum circuit using $X$ and $Z$ gates:
$$Q = e^{-i.t.(X \otimes Z \otimes Z \otimes Id \otimes X)}$$
that is introduced in Figure 5.

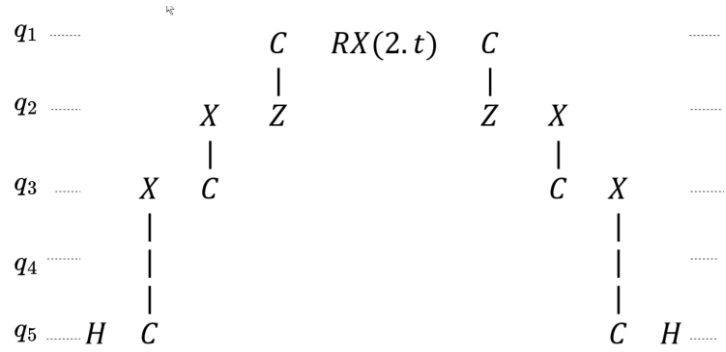

Figure 5. Quantum circuit for $e^{-i.t.(X \otimes Z \otimes Z \otimes Id \otimes X)}$

Step 2 Using $H.S$ and $S^\dagger.H$ we switch $X$ into $Y$ to generate $e^{i.t.X \otimes Z \otimes Z \otimes Id \otimes Y}$ and to obtain the quantum circuit of Figure 6.

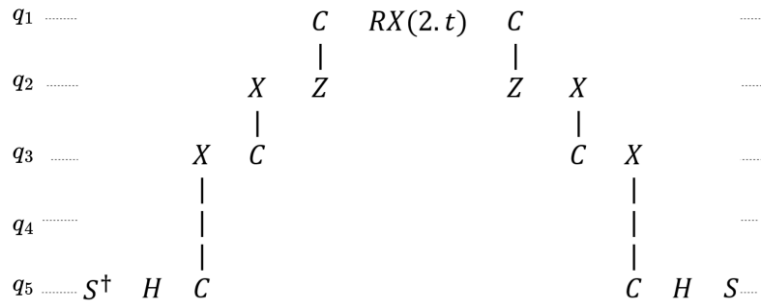

Figure 5. Quantum circuit for $e^{-i.t.(X \otimes Z \otimes Z \otimes Id \otimes Y)}$

## 3. Conclusion

In conclusion, the proposed algorithm for exponentiating quantum circuits provides a concise and comprehensible solution for defining the exponentiation of quantum circuits. The algorithm's justification and the calculations performed, and that rely on tensor computations only, ensuring clarity and avoiding computationally expensive matrix manipulations. Experimental results confirm its correctness in defining quantum circuits aligned with the Qiskit library's results. By proposing a simple algorithm, this work aims to make the concept more accessible to computer scientists, offering an algorithmic perspective rather than a purely mathematical one on creating quantum circuit exponentials.

**Acknowledgement**
Special thanks to the constructive remark of Julien Zylberman about some formulae.